\documentclass[prl,twocolumn]{revtex4}%
\usepackage{amsmath}
\usepackage{amsfonts}
\usepackage{amssymb}
\usepackage{graphicx}%
\begin{document}

\preprint{}
\title{Design, theory, and measurement of a polarization insensitive absorber for terahertz imaging}
\author{N.~I. Landy$^{1}$, C.~M. Bingham$^{1}$, T. Tyler$^{2}$, N. Jokerst$^{2}$, D.~R. Smith, and W.~J. Padilla$^{1}$}
\affiliation{$^{1}$Boston College, Department of Physics, 140
Commonwealth Ave., Chestnut Hill, MA 02467.}
\affiliation{$^{2}$Department of Electrical and Computer
Engineering, Duke University, Durham, NC 27708 USA.}

\begin{abstract}We present the theory, design, and realization of a
polarization-insensitive metamaterial absorber for terahertz
frequencies. We derive geometrical-independent conditions for
effective medium absorbers in general, and for resonant
metamaterials specifically. Our fabricated design reaches and
absorptivity of 65\% at 1.145 Thz.
\end{abstract}
\maketitle

\section{Introduction}

The advent of terahertz (THz) spectroscopy has ushered in a new
field of research as scientists seek to exploit the electromagnetic
signatures of materials in the THz range of the spectrum. Specifically, the
ability to perform imaging in the THz would have profound impact on
the areas of
security\cite{oliveiraSPIE03,zimdarsSPIE03,liuOpex03,barberJPCA05},
biology\cite{zhangPMB02,crowePTRSLA04}, and
chemistry\cite{jacobsenOptLett96,mittlemanAPL97}. However, imaging
in the terahertz is complicated by the a lack of easily accessible
electromagnetic responses from naturally occurring materials.
\cite{williamsRPP06,tonouchiNP07}

Electromagnetic metamaterials (MMs)\cite{pendry96,pendry99} are one
potential solution to overcoming this ``THz Gap"\cite{science04}.
The operational frequency of a given MM design is geometrically
scalable to other regimes of the electromagnetic spectrum. Various
MM structures have been shown to naturally couple to either the
electric and / or magnetic components of light in frequency ranges
from radio~\cite{wiltshire01}, microwave~\cite{smithPRL00},
mm-Wave~\cite{ekmel}, THz~\cite{science04}, MIR~\cite{soukoulis},
NIR~\cite{brueck}, to the near optical~\cite{optical}. A great deal
of effort has been exerted to create low-loss metamaterial devices
such as negative index (NI) structures \cite{smithPRL00} and
electromagnetic cloaks\cite{pendrySci06,schurigSci06}. However, the
substantial loss tangent that occurs at the center frequencies of
metamaterial resonances can be exploited as well. This resonant loss
phenomenon serves as the starting point for the investigation of MMs
as narrow-band absorbing elements for thermal imaging devices.

We present the theory, design, fabrication, and measurement of a
single-frequency metamaterial absorbing element. The simulated
design reaches a peak absorptivity of $95\%$ at 1.13 THz and the
fabricated structure reaches a measured absorptivity of $65\%$ at
1.145 THz. This absorptivity is comparable to existing MM absorber
designs \cite{landyPRL08,taoOpEx08}, but our design is also
polarization insensitive, which maximizes the absorption of light.

\section{Theory}

The absorptivity A($\omega$) of a given material is given by the
transmission T($\omega$) and reflectance R($\omega$) as
$A(\omega)=1-T(\omega)-R(\omega)$. In terms of the complex
transmissivity ($\tilde{t}$) and reflectivity ($\tilde{r}$), this
can be written as $A=1-|\tilde{t}(\omega)|^2-|\tilde{r}(\omega)|^2$.
Therefore, $A=1$ when $T=R=0$. In reference \cite{smithPRE05}, the
frequency-dependent transmissivity $\tilde{t}(\omega)$ was
determined to be dependent on the complex index of refraction
$\tilde{n}(\omega)=n_{1}+in_{2}$ and impedance
$\widetilde{Z}(\omega)=Z_{1}+iZ_{2}$ for a slab of length $d$ as:

\begin{equation}
t(\omega)^{-1}=\sin(\tilde{n}kd) -
\frac{i}{2}(\widetilde{Z}+\frac{1}{\widetilde{Z}})\cos(\tilde{n}kd).
\end{equation}

\noindent Where $k= \omega /c$ and $c$ is the speed of light in
vacuum. We use the convention where a subscripted 1 and 2 denote the
real and imaginary parts of a complex function, respectively.

As $\widetilde{Z}$ approaches unity (the free space value), the
reflectivity will drop to zero
and the transmissivity will be determined entirely by $\tilde{n}$:

\begin{equation}
t^{-1}=\cos(\tilde{n}kd) - i\sin(\tilde{n}kd).
\end{equation}

Upon substitution of the exponential forms this becomes:

\begin{equation}
t^{-1}=e^{-in_{1}kd}e^{n_{2}kd}.
\end{equation}

\noindent So the transmission ($T=|\tilde{t}|^2$) is

\begin{equation}
T=e^{-2n_{2}kd}. \label{matchedT}
\end{equation}

Therefore, as $n_{2}$ approaches infinity,

\begin{equation}
\lim_{n_{2}\to\infty}T=0.
\end{equation}

The combined dielectric and magnetic losses in the system are
characterized by $n_{2}$. Therefore the physical interpretation of
the above derivation is that, (in the absence of reflections), the
transmission of an electromagnetic wave with a given wavevector $k$
through a slab of given thickness $d$ is determined entirely by
losses in the slab. To create a very high absorber it is then
necessary for $\widetilde{Z}=1$ at a point where $n_{2}$ is large.

Precise control of $\tilde{n}$ and $\widetilde{Z}$ is necessary to
realize a high absorber. Electromagnetic MMs are prime candidates
for this task since they can be designed to couple to electric and
magnetic components of light. This enables precise tuning of the
complex, frequency-dependent permittivity $\tilde{\epsilon}(\omega)$
and permeability $\tilde{\mu}(\omega)$ of a MM slab. The index
$\tilde{n}$ and impedance $\tilde{Z}$ are in turn given by
$\tilde{n}(\omega)=\sqrt{\tilde{\epsilon}(\omega)\tilde{\mu}(\omega)}$
and
$\widetilde{Z}(\omega)=\sqrt{\tilde{\mu}(\omega)/{\tilde{\epsilon}(\omega)}}$.

MMs are typically highly resonant in $\tilde{\epsilon}$ and / or
$\tilde{\mu}$, where the relevant optical constants approximate the
form of a complex oscillator in frequency:

\begin{equation}
\epsilon(\omega),\mu(\omega) =\epsilon_{\infty},\mu_{\infty}+
\frac{F_{\epsilon,\mu}\omega^2}{\omega_{0\epsilon,\mu}^{2}-\omega^{2}-i\gamma\omega}
\label{pendrian}
\end{equation}
where $F$ is the oscillator strength, $\gamma$ is the damping,
$\omega_0$ is the center frequency of the oscillator, and
$\epsilon_\infty$,$\mu_\infty$ are high frequency contributions to
$\epsilon$,$\mu$. This form for an oscillator describes the
frequency response of metamaterials, and we term this a ``Pendrian"
after ref. \cite{pendry99}.

We consider a single frequency operation point, defined by $\omega_0$,
such that when $\tilde{\epsilon}(\omega_0)=\tilde{\mu}(\omega_0)$ then
$\widetilde{Z}(\omega_0)$=1 and $n_2(\omega_0)$ is maximized with a
value of:

\begin{equation}
n_{2}(\omega_{0}) = \frac{F^2}{(\gamma\omega_{0})^2}
\label{n2Fgamma}
\end{equation}

which, according to Eq. \ref{matchedT}, yields

\begin{equation}
A(\omega_0)=1-\mathrm{Exp}(-2\frac{F\omega_0^2d}{\gamma~c})
\end{equation}

For the more realistic case, when $\epsilon_\infty \neq\mu_\infty$,
then  $\widetilde{Z}(\omega_0) =
\sqrt{\mu_\infty/\epsilon_\infty}\neq 1$ and $A$ is no longer
determined solely by $n_2$. However, for large $n_2$, $T(\omega_0)$
remains low and $R(\omega_0)$ can be written approximately in terms
of $Z(\omega)$ \cite{smithPRE05} such that

\begin{equation}
R(\omega_0)=({\frac{Z(\omega_0)-1}{Z(\omega_0)+1}})^2
\end{equation}

Regardless of $\epsilon_\infty$ and $\mu_\infty$, an optimal
narrow-band absorber must maximize $F$ with respect to $\gamma$ in
both $\epsilon$ and $\mu$. $F$ is determined by the geometry,
filling factor, and conductivity of the two metallizations. $\gamma$
is determined by losses in the metalization and substrate. The
optimal absorber will then use the MM geometries with the maximum
possible filling factor. Furthermore, metallization and substrate
should be chosen to minimize losses at the operation frequency as
determined by Eq. \ref{n2Fgamma} above.

\section{Design}

From a theoretical viewpoint, it would seem simple to design a MM
that would provide such a response in $\epsilon$ and $\mu$. However,
there are several complications to the theoretical analysis
presented above due to the specific properties of MMs. For instance,
the periodicity inherent to most MMs contributes to spatial
dispersion, i.e. $\epsilon=\epsilon(\omega,\mathbf{k})$ and
$\mu=\mu(\omega,\mathbf{k})$.\cite{tretyakovbook} This spatial
dispersion causes the optical parameters to deviate from the pure
form of Eq. \ref{pendrian}. Spatial dispersion also causes
antiresonances in $\tilde{\mu}$ ($\tilde{\epsilon}$) due to
resonances in $\tilde{\epsilon}$ ($\tilde{\mu}$) \cite{liuPRE07}.
Additionally, conventional electric \cite{padillaPRB07,schurigAPL06}
and magnetic \cite{smithPRL00} MMs are highly coupled when they
share a center resonant frequency \cite{smithJAP06}. Typically,
electric metamaterials have higher order electric resonances, and
thus in general one does not necessarily have the condition that
$\epsilon_\infty=\mu_\infty$, leading to a nonzero reflectivity as
described above. In our particular case for the design of a
narrow-band high absorber, a solution to the latter problem was to
use two electric resonances; one to raise and lower the curve at the
operational frequency and one to couple to $\mu$ to create a region
of high absorption.

\begin{figure}
[ptb]
\begin{center}
\includegraphics[ width=3in,keepaspectratio=true
]%
{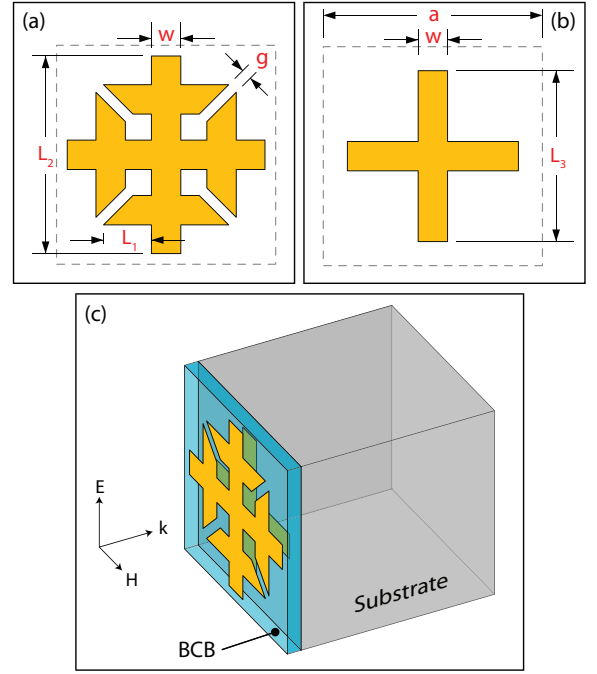}%
\caption{(color) (a) ERR (b) cross (c) combined ERR and cross. The
dimensions in microns are a $= 84$, L$_1 = 52.5$, L$_2 = 74$, L$_3 =
64$, w$ = 11$, and g$ = 4$. Axes indicate the wave polarization and
propagation direction. }
\label{fig1}%
\end{center}
\end{figure}

The electric responses were provided by a modified electrically
coupled ring resonator (ERR), shown schematically in
Fig.\ref{fig1}(a). The ERR chosen had fourfold rotational symmetry
about the propagation axis and was therefore polarization
insensitive \cite{padillaPRB07}. The lower-frequency electric
response used to tune the $\epsilon(\omega)$ curve was driven by the
LC loop in the ERR. The higher-frequency response used to couple to
$\epsilon(\omega)$ was created by the dipole-like interaction of the
metallizations in adjacent unit cells.

The magnetic response was created by combining the ERR with a cross
structure Fig.\ref{fig1}(b) in a parallel structure separated by a
layer of benzocyclobutane (BCB) Fig.\ref{fig1}(c). The magnetic
component \textbf{H} of a TEM wave coupled to the center stalks of
the two metallizations that were perpendicular to the propagation
vector, such that anti-parallel currents were driven. A
polarization-sensitive design based on similar parameters has been
shown at both THz \cite{taoOpEx08} and microwave\cite{landyPRL08}
frequencies.

Practical realization of metamaterials dictates the use of a support
substrate that complicates the theoretical optimization of the
structure.  This substrate is typically thick in the propagation
direction compared to the thickness of the metamaterial and the
lateral unit cell dimensions. This supporting structure introduces
ambiguity in the definition of the unit cell as well as significant
asymmetry in the propagation direction, both of which complicate the
definition of $Z_{eff}$ and $n_{eff}$.\cite{tretyakovPRB07} However,
some portion of the substrate must be included in simulation because
it has a measurable effect on the form of the MM resonances due to
the dielectric of the substrate, $\epsilon_{s}$
\cite{padillaJOSAB06}.

The effect of the substrate on the MM's resonances can be shown
directly in simulation. A parameter sweep of the substrate thickness
$d$ shows that the effect of the added dielectric on the MM
resonance saturates near a value of $d= d_{s}$. For $d
> d_{s}$, the only added effect is a factor
$\exp(i\sqrt{\epsilon_{s}}k(\omega)(d-d_{s})$ to
$\tilde{t}(\omega)$. As substrates are typically chosen for low loss
in the frequency range of interest, $|\tilde{t}(\omega)|$ is
virtually unchanged and the extended substrate only adds phase. The
appropriate unit cell boundary in the propagation direction is
therefore given by $d_s$ and the MM elements on the substrate. The
impedance mismatch at the MM-substrate and substrate-air boundaries
can then be incorporated into the homogenized effective medium as
these boundaries are within the unit cell.

\begin{figure}
[ptb]
\begin{center}
\includegraphics[ width=3in,keepaspectratio=true
]%
{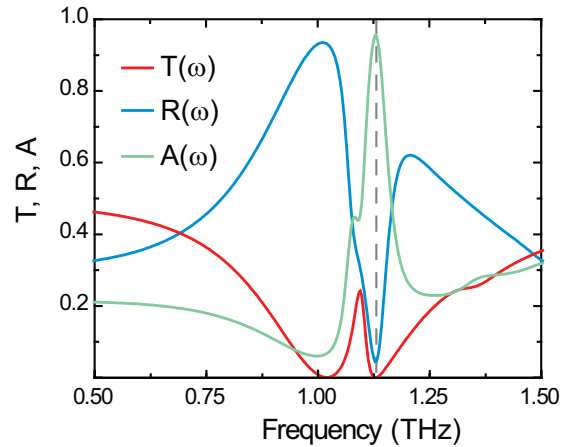}%
\caption{(color) Reflectance (blue), Transmission (red), and
Absorptivity (green) for the simulated absorber. The vertical dashed
line indicates the frequency of maximum absorptivity. }
\label{fig2}%
\end{center}
\end{figure}

\begin{figure}
[ptb]
\begin{center}
\includegraphics[ width=3.5in,keepaspectratio=true
]%
{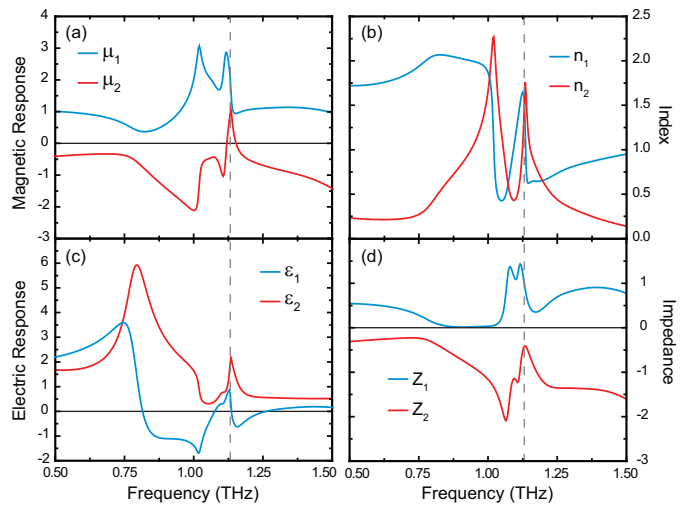}%
\caption{(color) Real (blue) and imaginary (red) components of
$\epsilon$ and $\mu$. Real (blue) and imaginary (red) components of
n and Z. The vertical dashed line indicates the frequency of maximum
absorptivity.}
\label{fig3}%
\end{center}
\end{figure}

The extracted optical constants for the metamaterial absorber shown
in Fig. \ref{fig1} are plotted in Fig. \ref{fig2}. The real and
imaginary components $\tilde{\epsilon}$ and $\tilde{\mu}$, as
plotted in Fig. \ref{fig2} (a),(c), are complicated by the effects
of spatial dispersion, as previously mentioned. However, various
features of the plots enable interpretation of the forms of these
curves. The lowest frequency feature in $\epsilon_2$ at $\omega=800$
GHz, shown in Fig. \ref{fig3}(c), is the conventional MM ELC
resonance. Notice this is accompanied by an antiresonance in
$\tilde{\mu}(\omega)$, which is characterized by a negative
imaginary component that peaks at the center frequency of
$\epsilon_1(\omega)$, defined by the cusp near $\omega=1.02$ THz,
and in accord with previous work\cite{liuPRE07}. A second electric
resonance appears at approximately $\omega=$1.125 THz due to the
cut-wire response of the cross. This is accompanied by an
antiresonance in $\tilde{\mu}(\omega)$ centered at the same
frequency. The resonance in $\tilde{\mu}(\omega)$ is weak relative
to the neighboring resonances in $\epsilon$, and thus difficult to
observe. However, $\mu_{2}$ has a distinct positive peak at 1.13
THz. This, combined with an approximately Pendrian curve in
$\mu_{1}$, indicates the presence of a magnetic response centered a
this frequency. There is also a small but distinct kink in
$\tilde{\epsilon}(\omega)$ due to the antiresonance caused by the
magnetic response.

Figure \ref{fig2}(b) shows the refractive index $\tilde{n}(\omega)$
and (d) impedance $\tilde{Z}(\omega)$ and demonstrates how close the
structure approximates an ideal absorber. At 1.13 THz the real
impedance is near unity, $Z_{1} \sim 1$, and the complex impedance
$Z_{2}$ is minimized, such that $R \sim 0$. As desired, the
imaginary index, $n_{2}$, is maximized near a value of $\sim 1.75$,
which minimizes T. This results in a peak absorptivity of $95\%$,
plotted as the green solid curve shown in Fig. \ref{fig2}.

\section{Fabrication \& Experiment}
The two layer metallization MM absorber sample was fabricated on a
high resistivity 1 mm thick silicon substrate.  A 3 um thick SiO2
layer was deposited on the Si substrate using plasma enhanced
chemical vapor deposition (PECVD). The bottom, cross-shaped
metallization (30 nm Ti/40 nm Pt/200 nm Au) was patterned using
standard negative lithography, metal evaporation, and metal
lift-off. The BCB dielectric (Cyclotene 3022-46, Dow) was deposited
using two consecutive spin-coat depositions and soft cures in a
vacuum oven, resulting in a final thickness of approximately six
microns. The top ELC metallization layer structure and patterning
used the same process as that of the cross. The two unit cells (one
for each layer) are shown as the insets to the bottom panel of
Fig.4.

The MM absorber sample was examined experimentally using a Fourier
Transform Infrared (FTIR) spectrometer. Polarized light from a
mercury arc-lamp was transmitted and reflected from both the sample
and a reference substrate, and then focused on the detector, a
liquid helium cooled Si-bolometer. Measurements of the sample for
both polarizations where characterized and we found no deviations
within experimental error. Measured $T(\omega)$ and $R(\omega)$ were
used to calculate the experimental $A(\omega)$.

Fabrication tolerances in the structures resulted in some deviation
from the theoretically optimized case, and are shown in Figure 4.
The corners of the ELC and cross structures are slightly rounded,
and the BCB thickness was 6 um rather than the optimal 5.8 um. All
of these factors were incorporated into the computer model and the
re-simulated transmission, reflection, and absorbtion for the
metamaterial absorber are shown in Figure 4, with the experimental
measurement of $T(\omega)$ and $R(\omega)$.  In comparison to the
theoretically optimized structure results shown in Figure 2, the
fabricated sample exhibits a shift in the MM resonances.  The first
transmissive minimum has shifted by 10GHz, while the second minimum
has shifted by 54GHz, both to higher frequencies. The minimum in
reflectance has shifted from 1.128 THz to 1.15 THz. This is expected
as electric responses are sensitive to metallic structure rounding
\cite{schurigSci06} and the thickness of the BCB layer partially
determines the magnetic coupling. However, this change in
geometrical parameters does not fully account for the disagreement
between Figures 2 and 4.

At 0.5 THz, $T(\omega)$ and $R(\omega)$ are well matched in
experiment and simulation, but the forms deviate as the curves
approach the MM resonances. Specifically, the experimental
reflectivity reaches a minimum of only $18\%$ compared to the
simulated value of $2\%$. Likewise, the first and second
transmissive minima reach values of $3\%$ and $3\%$, respectively,
as opposed to less than $0.1\%$ in simulation. Several mechanisms
may be behind this disparity.

The off-resonant agreement indicates that the discrepancies are
related to resonant forms of the constitutive parameters
$\tilde{\epsilon}$ and $\tilde{\mu}$. As indicated by Eq.
\ref{n2Fgamma} lowering $F$ or increasing $\gamma$ of the resonances
translates into a higher minimum in $T(\omega)$ as the peak value of
the effective loss ($n_2$) is decreased. This also changes the form
of $\tilde{Z}(\omega)$, and therefore the form of $R(\omega)$.

\begin{figure}
[ptb]
\begin{center}
\includegraphics[ width=3in,keepaspectratio=true
]%
{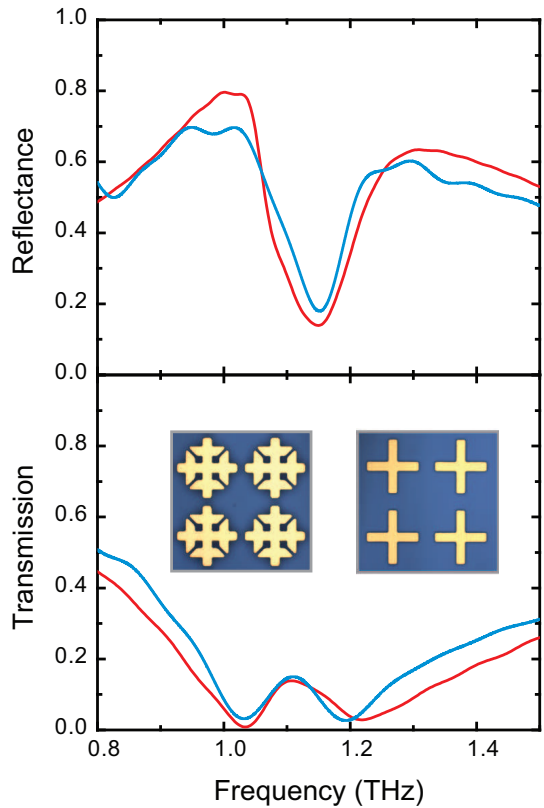}%
\caption{(color) Reflectance and transmission results for experiment
(blue) and simulation (red) }
\label{fig4}%
\end{center}
\end{figure}

For both the electric and magnetic resonances, $\gamma$ is primarily
determined by the loss in the BCB substrate between the ELC and
cross. Previous work has determined that dielectric losses are the
primary mechanism, and may be an order of magnitude greater than
Ohmic loss\cite{landyPRL08}. Also, because BCB is the primary
dielectric that tunes the capacitance of each structure, any
deviation from the nominal value of $\epsilon=$2.5 will lead to
variation in absorbance. Further, it is well known that the
dielectric value of many polymer compound have significant
dependence on frequency, especially within the THz
range\cite{taoAPL}. As previously mentioned, the ``strength of the
oscillator" $F$ is determined by the geometry, filling fraction, and
conductivity of the two metallizations. Therefore, the deviation of
experimental curves in Fig. \ref{fig4} may be caused by a
combination of increased loss in the BCB and lowered conductivity in
the metallizations.

\begin{figure}
[ptb]
\begin{center}
\includegraphics[ width=3in,keepaspectratio=true
]%
{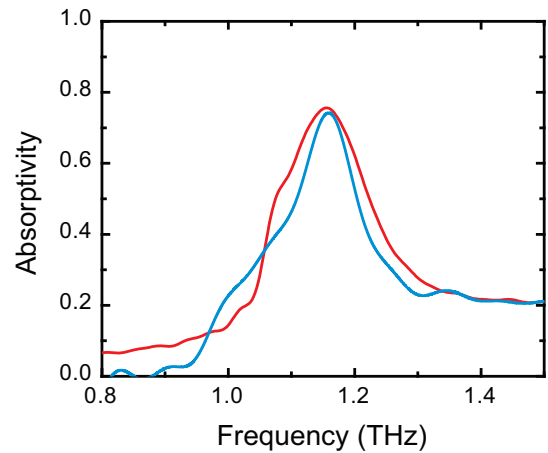}%
\caption{(color) Experimental (blue) and simulated (red)
Absorptivity curves. }
\label{fig5}%
\end{center}
\end{figure}

The best fit of simulation to experimental data is shown in Figures
4 and 5. The loss tangent of BCB was found to be approximately one
order of magnitude greater than the nominal value, while the
conductivity of gold was found to have decreased an order of
magnitude. As a consequence of the non-uniform shift of both the
magnetic and electric resonances, as well as increased damping due
to loss, $\tilde{n}$ and $\tilde{Z}$ have deviated from their
optimum values, resulting in a peak absorptivity of $74\%$ (Fig.
\ref{fig5}).

\section{Discussion}

We now discuss the potential use of the metamaterial absorber as a
thermal imager for the THz frequency range. Compared to existing THz
absorbers, our device is narrow band. This enables
spectrally-selective applications, such as in the detection of
explosives. However, unlike previous MM absorber
designs\cite{taoOpEx08}, our device is polarization insensitive.
This may be ideal for certain applications, as it maximizes
absorption for arbitrarily polarized or incoherent light. As a MM
device, our design is geometrically scalable to different frequency
ranges. This scalability is limited only by limitations in
fabrication and loss in constituent materials. The limitations of
the narrow band, resonant design could be overcome by using multiple
distinct unit cells\cite{bingham} or by incorporating tunable or
frequency agile metamaterial components \cite{chenNatPhot08}.

It should be noted that designs presented here are bianisotropic - a
result of asymmetry in the propagation direction - and belong to
Sh\"{o}nflies point group $C_4$.\cite{padilla07b} We have performed
simulations, (not shown), in order to elucidate the impact of the
bianisotropy on absorptive properties of the metamaterial. We
studied the cross polarization in transmission as a function of
frequency for the design shown in Fig. \ref{fig1}. Computer
simulations indicate that the cross polarization is small and
achieves a maximum of only 10$^{-4}$ over the frequency range of
interest. We also investigated the cross polarizations for designs
presented in Refs. \cite{landyPRL08,taoOpEx08} and found similar
results. Thus, for normal incidence radiation, the effect of
bianisotropy is negligible.

In conclusion, we have derived general conditions to create an
absorber based on effective medium theory, and for the specific case
of MM elements. We have shown that such a design can reach
absorptivities approaching unity within a narrow band. We have also
successfully implemented this approach with a THz frequency absorber
design. The theory presented here and the design specifically show
great promise to creating absorbers at any decade of frequency.

\end{document}